# Microstructure and formation mechanisms of δ-hydrides in variable grain size Zircaloy-4 studied by electron backscatter diffraction

Siyang Wang*, Finn Giuliani, T. Ben Britton

Department of Materials, Royal School of Mines, Imperial College London, London, SW7 2AZ, UK

*Corresponding author: siyang.wang15@imperial.ac.uk

## Abstract

Microstructure and crystallography of δ phase hydrides in as-received fine grain and 'blocky alpha' large grain Zircaloy-4 (average grain size ~11 μm and >200 μm, respectively) were examined using electron backscatter diffraction. Results suggest that the matrix-hydride orientation relationship is $\{0001\}_\alpha||\{111\}_\delta; <11\bar{2}0>_\alpha || <110>_\delta$ for all the cases studied. The habit plane of intragranular hydrides and some intergranular hydrides has been found to be $\{10\bar{1}7\}$ of the surrounding matrix. The morphology of intergranular hydrides can vary depending upon the angle between the grain boundary and the hydride habit plane. The misfit strain between α-Zr and δ-hydride is accommodated by high density of dislocations and twin structures in the hydrides, and a mechanism of twin formation in the hydrides has been proposed. The growth of hydrides across grain boundaries is achieved through an auto-catalytic manner similar to the growth pattern of intragranular hydrides. Easy collective shear along $<1\bar{1}00>$ makes it possible for hydride nucleation at any grain boundaries, while the process seems to favour grain boundaries with low (<40°) and high (>80°) *c*-axis misorientation angles. Moreover, the angle between the grain boundary and the adjacent basal planes does not influence the propensity for hydride nucleation.

## 1. Introduction

Zirconium alloys are used extensively as fuel cladding materials in water-based nuclear reactors due to their low neutron absorption cross section to mechanical strength ratio and relatively good corrosion resistance. Oxidation of the fuel cladding occurs in-service through its chemical reaction with the coolant water. Atomic hydrogen is thus released and can be readily absorbed by the cladding [1]. Supersaturation of hydrogen in zirconium leads to the formation of zirconium hydride precipitates [2–4], which can degrade the overall ductility and toughness of the cladding material [5–10]. Hydride embrittlement of Zr alloys through this process can result in the facture of the material through delayed hydride cracking (DHC) [4,11–13].

Four different Zr hydride phases have been observed so far, including trigonal ζ phase, face-centred tetragonal (FCT) γ phase (c/a>1), face-centred cubic (FCC) δ phase, and FCT ε phase (c/a<1) [14–16]. The most commonly observed hydride phase in fuel claddings is δ phase, therefore the formation mechanism and mechanical properties of δ phase hydride is of particular interest. It was reported that δ phase Zr hydrides tend to form preferentially at grain boundaries under slow cooling rates [17–20]. In Zircaloy-4, it has been found that the crystallographic orientation relationship (OR) between hexagonal close packed (HCP) α-Zr matrix and FCC δ-hydride is $\{0001\}_\alpha||\{111\}_\delta; <11\bar{2}0>_\alpha || <110>_\delta$ [19,21–23], and the habit planes of the macroscopic hydride packets are the $\{10\bar{1}7\}$ planes, which are close to the basal plane (~14.7°), of the surrounding α-Zr matrix [24,25]. The precipitation of δ-hydride in α-Zr results in a dilatational volumetric strain of 17.2% locally [26], and this misfit is accommodated by dislocations observed within both the hydride and the matrix [27,28].

In textured Zircaloy it was observed that the hydride stringers align perpendicular to the sheet or tube normal direction (ND) [22,29]. In thin wall cladding tubes, the radial hydrides are extremely detrimental to the ductility of fuel claddings as compared to circumferential hydrides, the texture of the cladding materials is controlled through processing such that hydrides form circumferentially during operation [30,31]. However, the reason why hydrides orient along specific texture components of Zircaloy is still not well understood. Kumar *et al*. [22] claimed that in textured Zircaloy-4 the intergranular hydrides form more easily in those grains with their basal planes oriented close to the grain boundaries, but no direct experimental evidence was provided. Theoretical calculations by Qin *et al*. [32] found that both grain boundary structure and energy may influence the propensity for hydride nucleation, and intergranular hydrides tend to nucleate at high energy grain boundaries with one of the grains having its basal plane nearly parallel to the grain boundary. Conflicting results on the preferred nucleation grain boundaries of hydrides in Zircaloy-2 have also been reported, including basal and prism tilt grain boundaries [33], grain boundaries making specific angles to the basal plane of the adjacent grains [34] and any grain boundaries [35,36].

Research efforts on hydride precipitation in Zr alloys so far are all carried out on as-received commercial Zircaloy cladding materials consist of relatively fine grains (around or less than 10 μm). Microstructural characterisation of δ phase hydrides in Zircaloy-4 using electron backscatter diffraction (EBSD) has significantly promoted the knowledge of the hydride distribution and hydride-matrix OR [21,22,27]. However, the focus of these works are mostly on the nature of the individual hydride packets instead of the interconnected hydride stringers which are of more significant industrial importance. The introduction of very large 'blocky alpha' grains (average grain size >200 μm) in Zircaloy-4 by Tong and Britton [37] opens up an opportunity for studying the grain size effect on hydride precipitation in the material and to significantly develop the work of Birch *et al*.[20]. The character of δ-hydrides in large grain Zircaloy-4 where the density of preferential nucleation sites are lower than the fine grain material may also provide useful industrial implications and new insights for the zirconium hydride study.

In the present work, EBSD is used to investigate the microstructural and crystallographic character of δ phase hydrides in both as-received fine grain (~11 μm) and large 'blocky alpha' grain (>200 μm) Zircaloy-4. The paper is organised as follows. Section 2 gives a description of the sample preparation methods and details about the EBSD experiments. Section 3 presents the EBSD results for the hydrides in large and fine grain Zircaloy-4 samples, and characters of the hydride packets are introduced individually in detail. In section 4, the results are discussed to deduce the formation mechanisms of hydrides, which are then used to assist the understanding of their crystallography and morphology.

## 2. Experiments

2.1. Materials and sample preparation

The material used in this study was cut from a rolled and recrystallised commercial Zircaloy-4 plate with a nominal chemical composition of Zr-1.5%Sn-0.2%Fe-0.1%Cr (wt%) [38]. The average grain size of the as-received material is ~11 μm, and the texture is a typical split basal texture with basal poles oriented ±30° away from the plate ND towards the plate transverse direction (TD). An annealing at 800 °C for 14 days was used for the production of samples consisting of large 'blocky alpha' grains [37]. Hydrogen charging of the samples was carried out electrochemically in a solution of 1.5 wt% sulfuric acid using a current density of 2 kA/m$^2$ at 65 °C for 24 hours, resulting in a hydrogen content of ~139 wt-ppm. After hydrogen charging, the samples were annealed at 400 °C (i.e. above the hydride solvus line for these alloys) for 5 hours to homogenise the hydrogen distribution, followed by a controlled slow furnace cooling of 0.5 °C/min to room temperature to promote the formation of δ phase hydrides. The samples were then mechanically polished with colloidal silica and electropolished in a solution of 10 vol% perchloric acid in methanol at −40 °C for 90s under an applied voltage of 30 V to remove the surface stress layer and improve the quality of the subsequent EBSD characterisation.

2.2. EBSD characterisation

EBSD scanning was conducted on an FEI Quanta 650 scanning electron microscope (SEM) equipped with a Bruker eFlashHR (v2) EBSD camera, using a beam accelerating voltage of 20 kV and probe current of ~10 nA. Collection and analysis of the EBSD data were carried out using Bruker ESPRIT 2.1 software. Depending on the experiment, EBSD patterns were binned from a native resolution of 1600×1200 pixels to 320×240 pixels, with an exposure time of 22ms.

## 3. Results

3.1. General hydride distribution

Inverse pole figure (IPF) maps of the as-received fine grain and 'blocky alpha' large grain samples before hydrogen charging are given in Figure 1(a) and (b), respectively. Figure 1(c) and (d) show the distribution of hydrides within the fine and large grain Zircaloy-4, respectively.

In the fine grain material, most of the hydrides are found at grain boundaries. Macroscopically, the interconnected hydride stringers align perpendicular to ND, in agreement with the observations in the literature [22,29]. In the large grain sample, three types of hydrides are observed which can be described in terms of their different nucleation sites:

- (1) Intergranular hydrides are those observed at the matrix grain boundaries. It is obvious that the proportion of intragranular hydrides is higher in the large grain sample than in the fine grain sample, which is likely a result of the smaller number/area of grain boundaries per unit volume in the large grain material (similar to observations of Birch et al. [20]).

(2) Twin boundary hydrides can be seen at the matrix twin boundaries. Twin boundaries are preferential nucleation sites for hydride formation in commercial fine grain Zircaloy-4 [39], and this has also been found to be true in the 'blocky alpha' microstructure (Figure 1(d)).

(3) Intragranular hydrides are observed inside the 'blocky alpha' grains.

The intergranular and twin boundary hydrides are mostly thicker than the intragranular hydrides, and a depletion of intragranular hydrides can be seen in the regions near the grain boundaries and the twin boundaries.

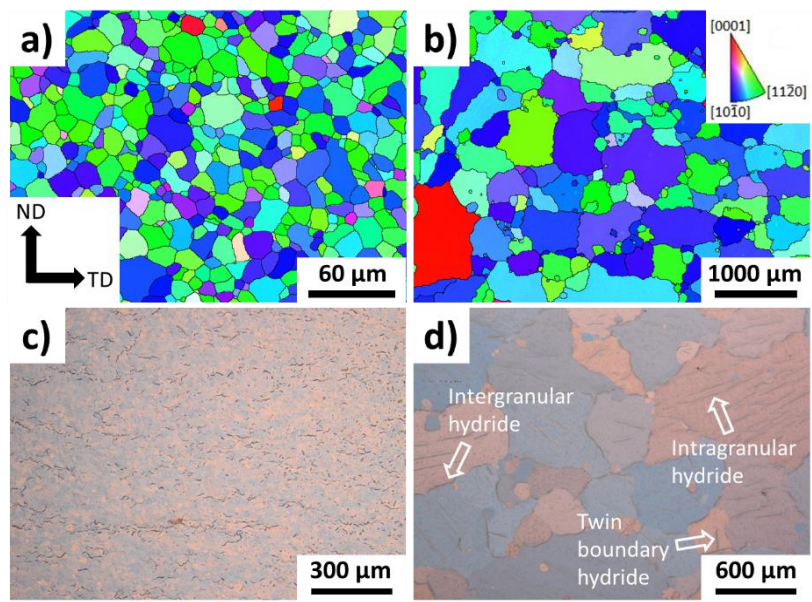

Figure 1 Inverse pole figure - rolling direction (IPF-RD) map of the (a) as-received fine grain and (b) 'blocky alpha' large grain Zircaloy-4; Polarised light optical micrograph of hydrided (c) fine grain and (d) large grain Zircaloy-4. The directions of the texture components are the same for the four images, as labelled in (a).

## 3.2. Hydrides in 'blocky alpha' large grain Zircaloy-4

### 3.2.1. Intragranular hydride

Figure 2 shows the crystal orientation map and the pole figures of a local area within a 'blocky alpha' grain containing an intragranular hydride packet. The blue region in the map corresponds to the hydride packet, while the red region corresponds to the zirconium matrix. In the hydride packet there exists two hydride grains with different crystal orientations (shown within the pole figures).

Analysis of the pole figures enables analysis of the orientation relationships between the α-zirconium matrix and the two observed hydrides. Comparing the pole figures of the δ hydrides with α-Zr matrix, both hydride grains follow the $\{0001\}_\alpha || \{111\}_\delta; <11\bar{2}0>_\alpha || <110>_\delta$ OR with the surrounding matrix.

For the hydride pole figures, the poles within the red circles are indexed points inside smaller hydride (contained in the white box within the map) and those highlighted with red squares correspond to the points outside the white box and for the larger hydride area. The pole figures of the δ-hydride suggest that the two hydride grains share one of the $\{111\}$ planes and three of the $<110>$ directions, indicating that they are $\{111\}<11\bar{2}>$ twins, and the twinning axis is the <c> axis of the zirconium matrix. In the pole figures, the projections of the hydrides are more dispersive than those of the matrix, implying that the degree of crystal misorientation in the hydride phase is higher than that in the matrix.

Further to this crystallographic analysis, spatial analysis of the hydride packet with the orientation of the matrix confirms that the trace of the habit plane of this hydride packet is nearly parallel to one of the $\{10\bar{1}7\}$ planes but not to the basal plane of the matrix surrounding it. Furthermore, the hydride twin interface is colinear with the basal plane of the α-zirconium matrix.

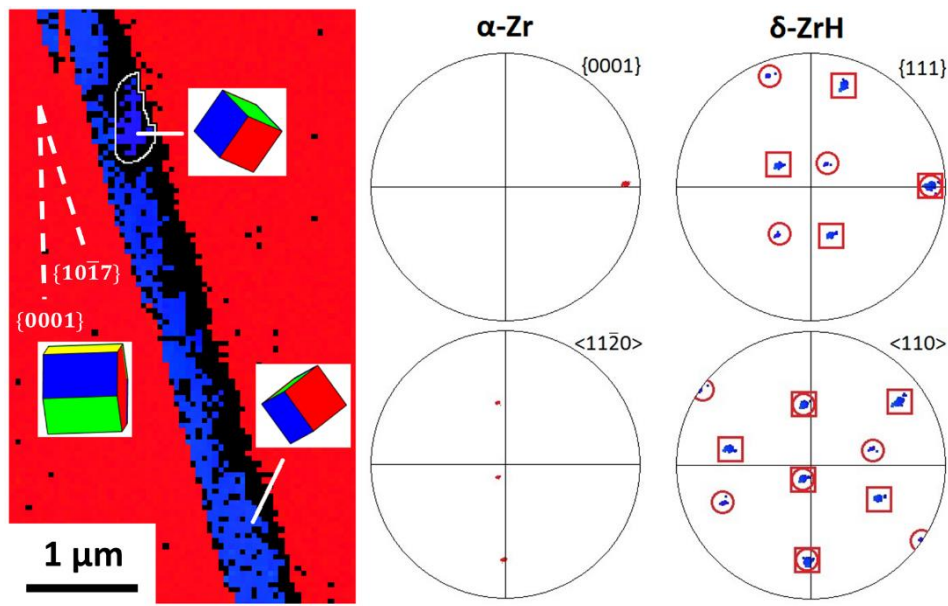

Figure 2 Crystal orientation map and pole figures of an intragranular hydride packet in Zircaloy-4. In the pole figures, the poles within the red circles are indexed points inside smaller hydride (contained in the white box within the map) and those highlighted with red squares correspond to the points outside the white box and for the larger hydride area.

### 3.2.2. Twin boundary hydride

EBSD results of a typical twin boundary hydride packet are shown in Figure 3. The crystal orientation map clearly shows that two crystal orientations of the α-Zr matrix can be observed within the field-of-view, and their OR indicates that they are $\{10\bar{1}2\}<\bar{1}011>$ (T1) twins.

The crystal orientation map and the pole figures show the presence of two pink hydrides, both sitting on the parent grain side at the twin boundaries. These hydrides follow the $\{0001\}_\alpha||\{111\}_\delta; <11\bar{2}0>_\alpha || <110>_\delta$ OR with the blue-coloured parent α-Zr grain. Similarly, the two purple hydrides sitting on the twin side at the twin boundaries, follow the same OR with the green-coloured matrix twin.

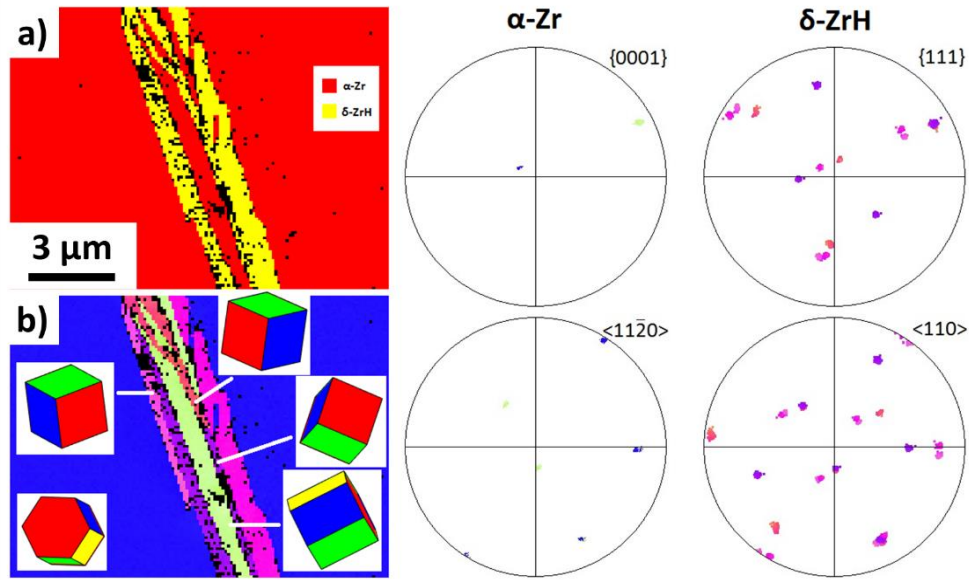

Figure 3 (a) phase map, (b) crystal orientation map and pole figures of a twin boundary hydride packet in Zircaloy-4. Colouring in the pole figure corresponds to the map colouring in (b). Hydride packet contains twin structure.

There is third hydride orientation, coloured red in the crystal orientation map, which resides inside the green-coloured α-Zr grain. This hydride also follows the $\{0001\}_\alpha||\{111\}_\delta; <11\bar{2}0>_\alpha || <110>_\delta$ OR with the surrounding matrix, and is a $\{111\}<11\bar{2}>$ twin to the purple-coloured hydrides. The relatively dispersive hydride projections in the pole figures imply a higher internal lattice rotation gradients likely due to a high density of geometrically necessary dislocation within the hydrides, and this is greater than in the matrix. The macroscopic habit plane of this hydride packet follows the $\{10\bar{1}2\}$ matrix twin plane.

To summaries this complicated structure, we observe hydrides at a T1 twin. The hydrides have an OR with the neighbouring zirconium twin or parent, and therefore also have an orientation relationship with each other. The third hydride has an OR with the neighbour α-zirconium and is a twin with its neighbour hydride.

### 3.2.3. Intergranular hydride

Observations of the large grain sample using SEM reveal that there are two different types of intergranular hydrides in the material, as shown in Figure 4. Some of the intergranular hydrides are along the grain boundaries (*type-a*) while the others are not (*type-b*), and usually the former appear to be thicker than the latter in morphology. The presence of these two types of intergranular hydrides has also been reported in fine grain Zircaloy-4 [22,27,40], but not analysed in detail. EBSD analyses were then carried out on these two different types of intergranular hydrides to study the origins of their different morphologies.

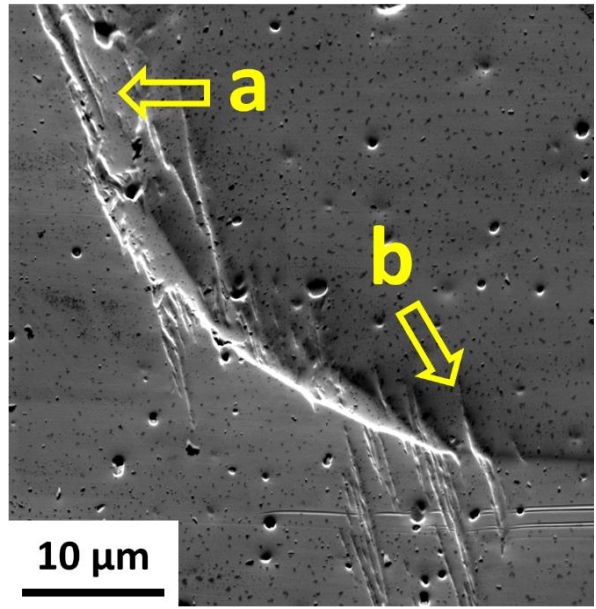

Figure 4  SEM micrograph showing two different types of intergranular hydrides: hydrides along the grain boundaries (*type-a*) and hydrides nucleated at the grain boundaries and grew into the grain interior (*type-b*).

An example of the *type-a* intergranular hydride is given in Figure 6, where the hydride packet (purple-coloured) can be observed to form along the grain boundary. The $\{0001\}_\alpha || \{111\}_\delta; <11\bar{2}0>_\alpha || <110>_\delta$ OR is satisfied between the hydride packet and the orange-coloured matrix grain on the right side of the grain boundary. From this local OR, it is therefore inferred that the hydride packet is transformed from the orange-coloured matrix grain during the cooling process. The pole figures also suggest that a higher degree of orientation gradient is present in the hydride than in the α-Zr matrix, similar to the previous observations in the intragranular and the twin boundary hydrides.

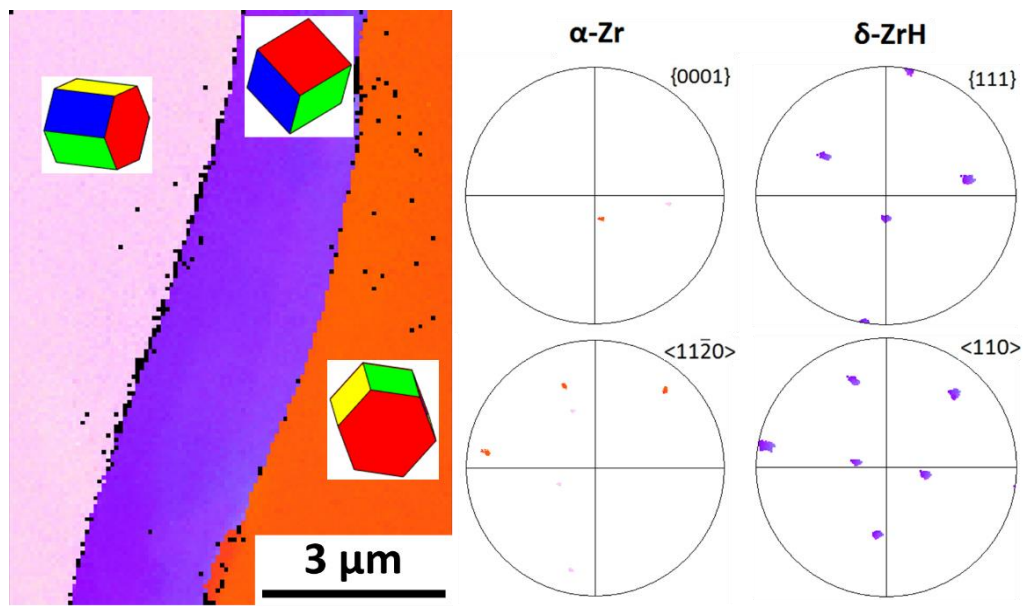

Figure 5  Crystal orientation map and pole figures of a *type-a* intergranular hydride packet along the grain boundary in Zircaloy-4.

A second example of *type-a* intergranular hydride is shown in Figure 6. However in this case multiple crystal orientations can be seen in the hydride on both sides of the grain boundary.

The zirconium grain boundary has a slight misorientation of the <c> axis and a significant twist about the <c> axis (mismatching the <a> type directions).

The three hydride orientations can be matched with the $\{0001\}_\alpha||\{111\}_\delta; <11\bar{2}0>_\alpha || <110>_\delta$ OR with their adjacent matrix grains, as can be seen in the pole figures. The blue- and orange-coloured hydrides are $\{111\}<11\bar{2}>$ twins to each other. There are two orange hydride grains, one on the top right in the crystal orientation map and another inside the blue hydride grain.

Additionally to the two matrix α-zirconium grains, there is a small region of α-zirconium contained within the hydride packet and at the interface of two of the orange and the blue hydrides. This small grain of zirconium has a slightly different crystal orientation to the bulk grain on the right, which is also noticeable in the pole figures, where the projections from the small matrix region between the hydrides are highlighted with red squares.

Again, the orientation gradients within the hydrides are more significant than that in the matrix, as suggested by spread of orientations within the pole figures.

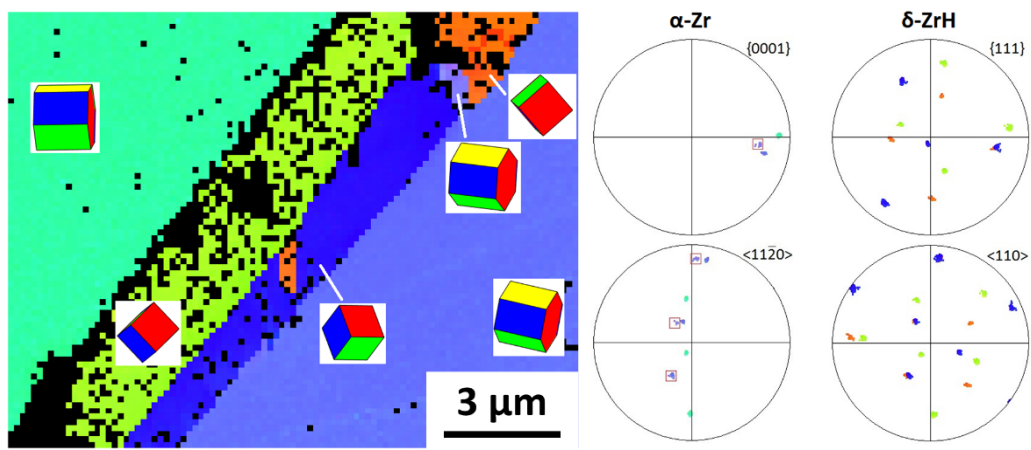

Figure 6 Crystal orientation map and pole figures of an intergranular hydride packet along the grain boundary in Zircaloy-4. Hydride packet contains twin structure.

Figure 7 shows the local crystal orientation map of another *type-a* intergranular hydride packet along the grain boundary. The yellow- and light green-coloured regions on both sides of the grain boundary correspond to the two matrix grains, while four hydride crystal orientations (red, pink, orange and purple) are present as well. The $\{0001\}_\alpha||\{111\}_\delta; <11\bar{2}0>_\alpha || <110>_\delta$ OR is satisfied between the yellow-coloured matrix and the two hydride orientations on the same side of the grain boundary (red and pink), as well as between the light green-coloured matrix and the other two hydride orientations (orange and purple) on the other side of the grain boundary. As before, populations of hydrides attached to the same matrix are $\{111\}<11\bar{2}>$ twin pairs, such as the red- & pink-coloured hydrides and the orange- & purple- coloured hydrides. Similar to prior hydrides, sharper projections of the α-Zr phase than the δ hydride phase suggest a lower degree of misorientation within the matrix than in the hydride. As can be observed clearly in the crystal orientation map, the twin boundaries in the hydrides on both sides of the grain boundary coincide at the grain boundary.

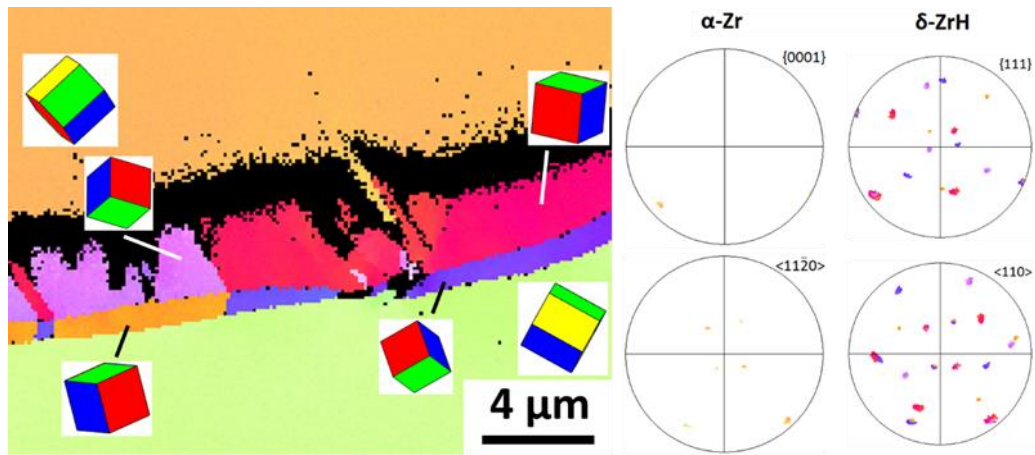

Figure 7  Crystal orientation map and pole figures of an intergranular hydride packet along the grain boundary in Zircaloy-4. Hydride packet contains twin structure and the twin boundaries coincide at the matrix grain boundary.

Next, a *type-b* intergranular hydride is introduced in Figure 8. This hydride packet is attached to a grain boundary and the macroscopic habit plane is nearly perpendicular to the grain boundary (the grain boundary lies horizontally in the crystal orientation map as labelled in the figure). The two dark blue hydrides in the map are twins to each other. The hydrides and matrix adhere to the $\{0001\}_\alpha||\{111\}_\delta; <11\bar{2}0>_\alpha\ ||\ <110>_\delta$ OR. The habit planes of the hydride packets in both grains are found to align more closely with one of the $\{10\bar{1}7\}$ planes than to the basal planes of the grains that they grew into respectively. As per pervious observations, a higher degree of misorientation can be seen in the hydride phase than in the surrounding matrix.

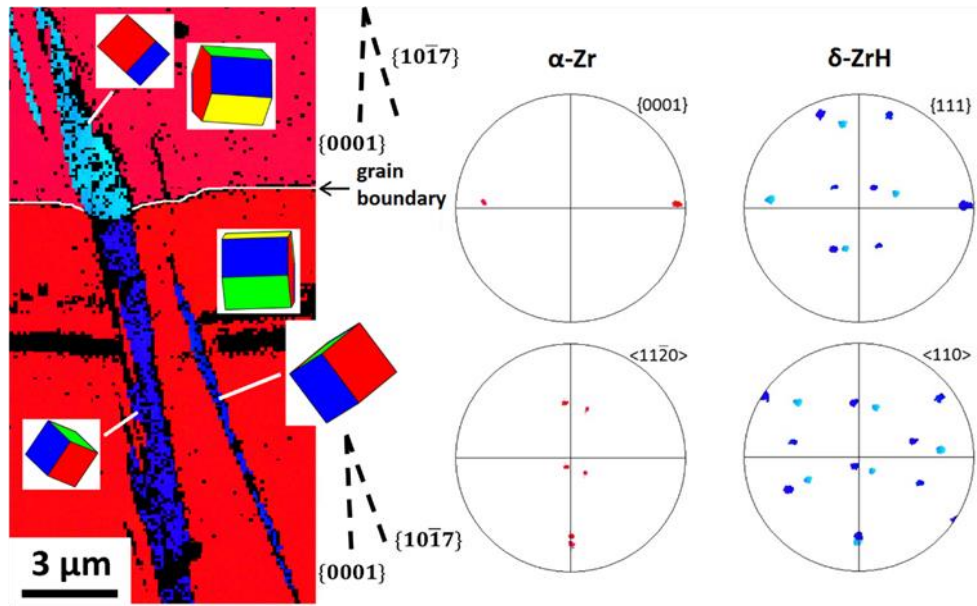

Figure 8 Crystal orientation map and pole figures of an intergranular hydride packet nearly perpendicular the grain boundary in Zircaloy-4. The macroscopic habit planes of the hydrides are close to the $\{10\bar{1}7\}$ planes of the surrounding matrix grains.

The $\{0001\}$ pole figure of the α-Zr phase also reveals that the two matrix grains involved have similar basal pole orientations. Furthermore, it seems that the precipitation of this intergranular hydride has led to a local perturbation of the position of the grain boundary, as can be seen in the crystal orientation map. This implies that there is a sympathetic nucleation and growth of these hydrides associated with this particular boundary.

### 3.3. Hydrides in as-received fine grain Zircaloy-4

A hydride-containing region of ~100 μm × 200 μm in the as-received fine grain material (Figure 1(c)) was studied using EBSD, and the results are shown in Figure 9. Generally, most of the hydrides observed are intergranular hydrides, while a relatively small fraction of the hydrides are thin intragranular hydride plates nucleated inside the grains. The macroscopic habit planes of the intragranular hydrides are found to be $\{10\bar{1}7\}$. Similar to the hydrides in the 'blocky alpha' microstructure, some hydride packets have two different crystal orientations internally which are $\{111\}<11\bar{2}>$ twins to each other.

The intergranular hydrides at vertical grain boundaries in the map (perpendicular to ND) are mostly *type-a* intergranular hydrides and appear to be thicker than those at horizontal grain boundaries which mostly belong to *type-b*.

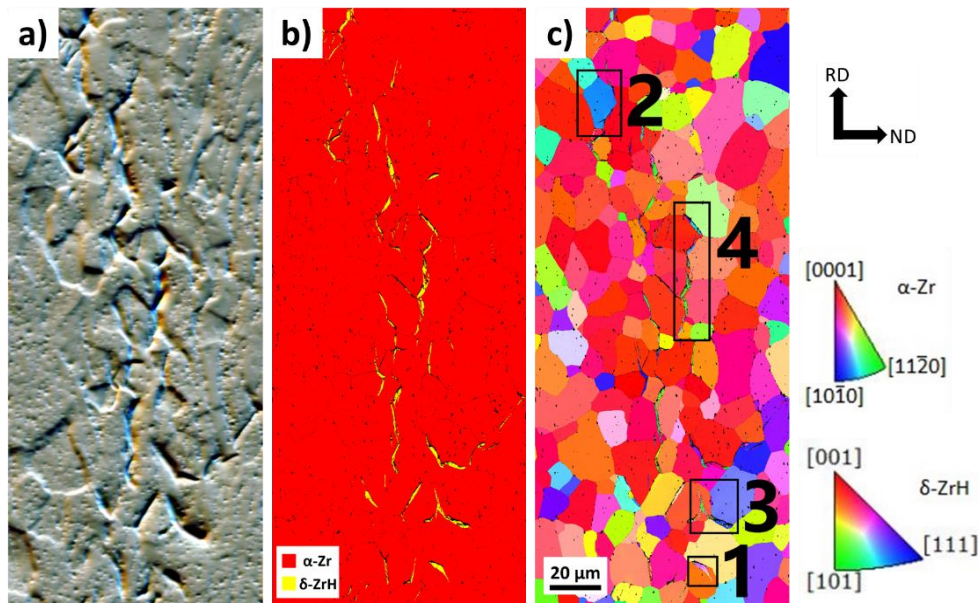

Figure 9 (a) Forescatter electron image, (b) phase map, (c) IPF-ND map of a hydride stringer in fine grain Zircaloy-4.

From Figure 9, four regions of interest in the map were picked out and the magnified local IPF-ND maps are given in Figure 10.

The hydride packet in region 1 (Figure 10(a)) is an intergranular hydride, containing two crystal orientations which are twins to each other. A demonstrable orientation gradient can be seen in the hydride packet, likely developed progressively during its growth process.

The IPF-ND map of region 2 (Figure 10(b)) shows no clear evidence of hydride formation in the blue grain, however multiple intergranular hydride packets can be observed in the neighbouring grains. Figure 10(e) shows the misorientation to reference point (the red cross at the centre of the grain) map and it is evident that local misorientations of up to 3° in this grain are present adjacent to the intergranular hydrides in the neighbouring grains. For those grain boundaries that are hydride-free, no significant local misorientation can be seen.

Region 3 (Figure 10(c)) contains two hydride packets, both having internal twin structures, in two adjacent grains. Changes in twin variants can be seen in the hydride packet in the orange-coloured matrix grain to have occurred repeatedly upon growth. Since these two hydride packets are interconnected at the grain boundary, it is inferred that the formation processes of them are related to each other.

A long hydride stringer extending across several grains can be observed in region 4 (Figure 10(d)). The hydride stringer consists of multiple intergranular hydride packets. These hydride packets, though formed in different grains, are connected with the hydrides in the adjacent grains at the grain boundaries. Lots of internal twin structures are observable within the hydride packets.

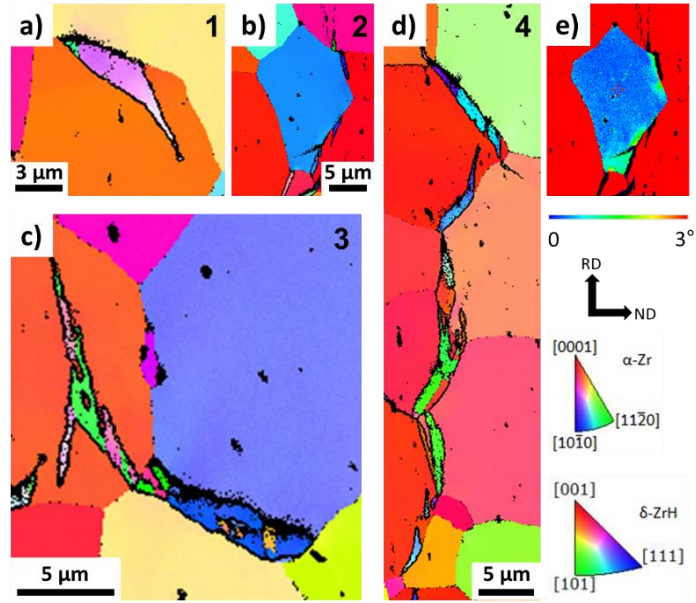

Figure 10 Magnified IPF-ND maps of region (a)1, (b)2, (c)3, (d)4 in Figure 9, (e) crystal misorientation to reference point (red cross) map of (b). The hydride-matrix phase boundaries are highlighted with black lines.

## 4. Discussion

For all the δ-hydrides studied in this work, including different types of hydrides in both the 'blocky alpha' large grain and the as-received fine grain Zircaloy-4, $\{0001\}_\alpha||\{111\}_\delta; <11\bar{2}0>_\alpha || <110>_\delta$ is the only observed OR between the δ-hydrides and the α-Zr matrix where the hydrides form. This indicates that the OR between these two phases is independent of the nucleation site of the hydrides and the grain size of the matrix material.

Generally, hydride packets were found to have a higher degree of crystal misorientation than the surrounding α-Zr matrix. This is indicative of a higher dislocation density in the hydrides than in the matrix, which is in agreement with results obtained from X-ray diffraction [28] and neutron diffraction [27] experiments, as well as electron channelling contrast imaging (ECCI) observations [23]. The dislocations produced during the transformation processes are presumably associated with the accommodation of the misfit strain between the matrix and the hydride.

The precipitation of δ-hydrides in α-Zr is achieved through an HCP to FCC phase transformation. This transformation, as proposed by Carpenter [41], may occur through the glide of $\frac{a}{3}<1\bar{1}00>$ partial dislocations on alternate {0001} planes, leaving behind stacking faults and thus changing the stacking sequence of the atomic layers. A schematic diagram of the transformation is reproduced in Figure 11.

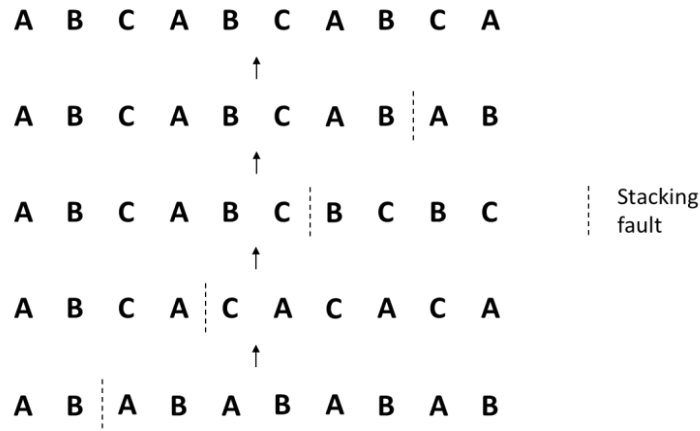

Figure 11  Schematic illustration of the HCP(α-Zr) to FCC(δ-hydride) phase transformation, adapted from [41].

Figure 12 illustrates this transformation with the motion of the atoms. It can be seen from the bottom five atomic layers that as the transformation progresses, strain is built up at the hydride-matrix interface and gradually increases. As a result, the gliding of partial dislocations with an opposite Burgers vector ahead of the growth front may become more energetically favourable which can then mitigate the strain progressively. This, as shown in the figure, can lead to the formation of twins in the hydride, which explains the routine observation of twin structures within the hydrides. As such, the repeated strain build-up and mitigation may cause the alternate formation of twin variants upon hydride growth as observed in Figure 10(c).

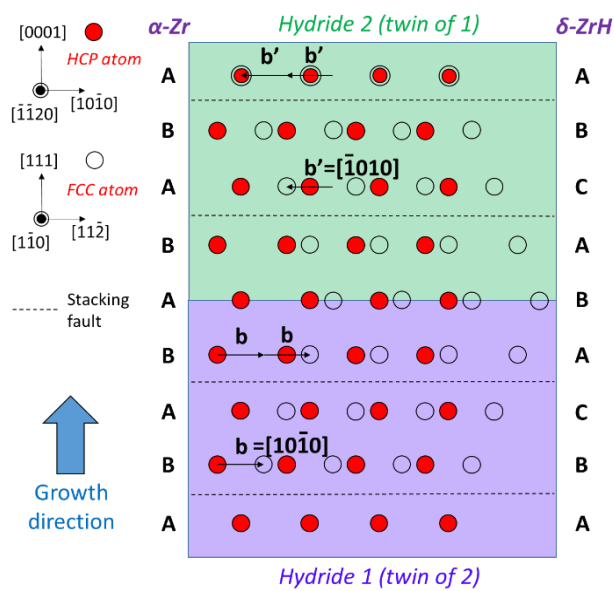

Figure 12  Schematic illustration of twin formation in δ-hydrides.

In the 'blocky alpha' large grain sample, some intergranular hydrides sit along the grain boundaries while the others seem to have nucleated at the grain boundaries and then grew into the grain interior, with their habit planes close to the $\{10\bar{1}7\}$ plane of the matrixes around them, similar to the habit planes of intragranular hydrides.

These two types of intergranular hydrides are also observed at grain boundaries in the fine grain material. This implies that after the initial nucleation event at the grain boundaries, the subsequent nucleation sites for further growth can vary depending on the local configuration of the chemical potential distribution.

According to the auto-catalytic nucleation mechanism of intragranular hydrides [18,42], upon hydride growth chemical potential wells are dynamically built up at the growth front which controls the path for further nucleation and growth. However, in the case of intergranular hydrides, the growth pattern is controlled by the competition between the chemical potential well produced by the hydride that has already formed and the initial chemical potential well at the grain boundary. Their relative hydrogen affinities likely determine the growth path of the intergranular hydrides.

The formation of *type-a* intergranular hydrides is presumably through significant amount of nucleation events at the grain boundaries and the thickening of the hydride packets afterwards. The formation processes of *type-b* intergranular hydrides, on the other hand, are dominated by the chemical potential well at the hydride growth front, which leads to a growth pattern similar to that of the intragranular hydrides and hydride depletion nearby at the grain boundaries (Figure 8).

The precipitation of δ-hydrides may result in orientation gradients within the matrix ahead of the growth front, and this can happen not only in the grain where the hydride formed (Figure 6) but also in the adjacent grain (Figure 10(b, e)). The orientation gradients observed are presumably due to the presence of dislocations produced to accommodate the misfit strain. Intergranular hydrides in the fine grain material are often interconnected at the grain boundaries (Figure 10(c, d)), implying their formation processes are not irrelevant.

Combining these two observations, a formation mechanism for the interconnected hydride stringers can be proposed as follows. During the moderately slow cooling process, hydride nucleation occurs preferentially at grain boundaries. The intergranular hydrides formed initially can produce dislocations in the neighbouring grains due to misfit strain ahead of the grown hydrides. This leads to the formation of chemical potential wells in the dislocated areas. Solute hydrogen atoms nearby then tend to diffuse towards these chemical potential wells. When local supersaturations of hydrogen are reached, new hydride nucleation events occur and these nuclei may grow out. Hence, similar to the growth of intragranular hydrides through repeated auto-catalytic nucleation, the growth of hydrides across grain boundaries is also achieved *via* an auto-catalytic mechanism. This creates chains of hydrides as stringers. The growth direction of these stringers is therefore dependent on the orientation of the repeated successful nucleation of hydrides as the stringers grow in the network.

Moreover, the fact that the twin boundaries in the hydrides on both sides of the grain boundary coincide at the grain boundary, as shown in Figure 7, suggests further that the formation of intergranular hydrides in adjacent grains cannot be considered as independent processes. As HCP structure is highly anisotropic, the dislocations produced by an intergranular hydride at the other side of the grain boundary may make it easier for the matrix to locally shear and transform into one hydride orientation than its twin orientation, which results in the observed intergranular hydride pairs at the grain boundaries.

It is therefore speculated that the characters of the grain boundaries may have an impact on the propensity for hydride formation, and the alignment of the shear planes and shear directions responsible for the transformation, $\{0001\}$ and $<1\bar{1}00>$, are likely of particular importance. Their respective effects on the hydride formation are thus extracted below.

Analysis of the *c*-axis ({0001} basal plane normal) misorientation angles ($\theta_c$) of all the grain boundaries where intergranular hydrides have formed is shown in Figure 9. The $\theta_c$ values for all the grain boundaries in Figure 1 were also calculated in order to reveal the intrinsic *c*-axis misorientation distribution for the textured plate. The two sets of data are plotted in Figure 13 direct comparison.

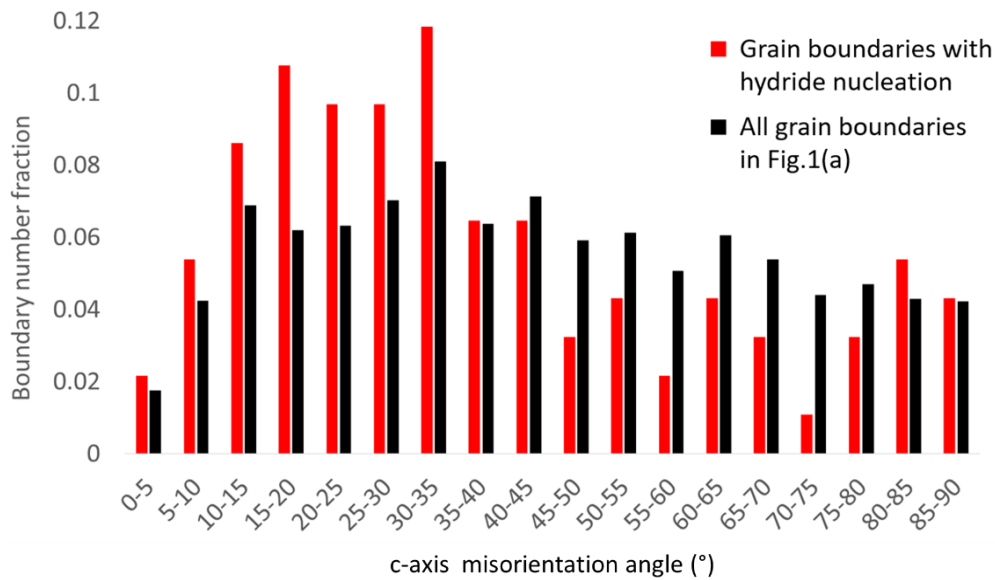

Figure 13  EBSD-based distribution of grain boundaries with c-axis misorientation angle.

It can be seen that the precipitation of intergranular hydrides seems to favour both the low $\theta_c$ (<40°) and the high $\theta_c$ (>80°) grain boundaries, while hydride formation at grain boundaries with $\theta_c$ between 40° and 80° is observed to be less frequent than the population of boundaries within these samples.

For low $\theta_c$ grain boundaries, the collective shear on the basal planes in both grains can take place relatively readily which may make it easy for hydride nucleation to occur (the case in Figure 8, for example). For high $\theta_c$ grain boundaries, however, the observed propensity for hydride nucleation may be attributed to high grain boundary energy.

It is also noticed that hydride nucleation is possible at grain boundaries with any $\theta_c$ value, which implies that the misorientation angle of the $<1\bar{1}00>$ shear directions available ($\theta_a$) may also have an influence on hydride nucleation.

Therefore, to reveal the $<1\bar{1}00>$ misorientation angles for different types of grain boundaries, we take an HCP crystal with Euler angles of $[\varphi_1,\phi,\varphi_2]=[0,0,30°]$ as a reference (crystal A). We then misorient a second crystal (crystal B) by varying its orientation through manipulating its Euler angles $\phi$ (i.e. the *c*-axis mismatch) and $\varphi_1$ which is the misorientation around the *c*-axis. The $\varphi_2$ for crystal B is forced to 30° for all cases. Figure 14 shows the change in the cosine of $\theta_a$ between these two crystals with the $\phi$ and $\varphi_2$ of crystal B. The conditions included in the figure are representative of the $\theta_a$ between any two randomly orientated HCP crystals due to the symmetry of the HCP crystal structure. It is evident that the cos $\theta_a$ values for all types of grain boundaries are high and the lowest possible value is 0.75, corresponding to a $\theta_a$ of 41.4°. This means that for any grain boundary the $<1\bar{1}00>$ misorientation angle is always low, allowing for easy collective shear and making it possible for hydride nucleation at any grain boundaries as observed in Figure 13.

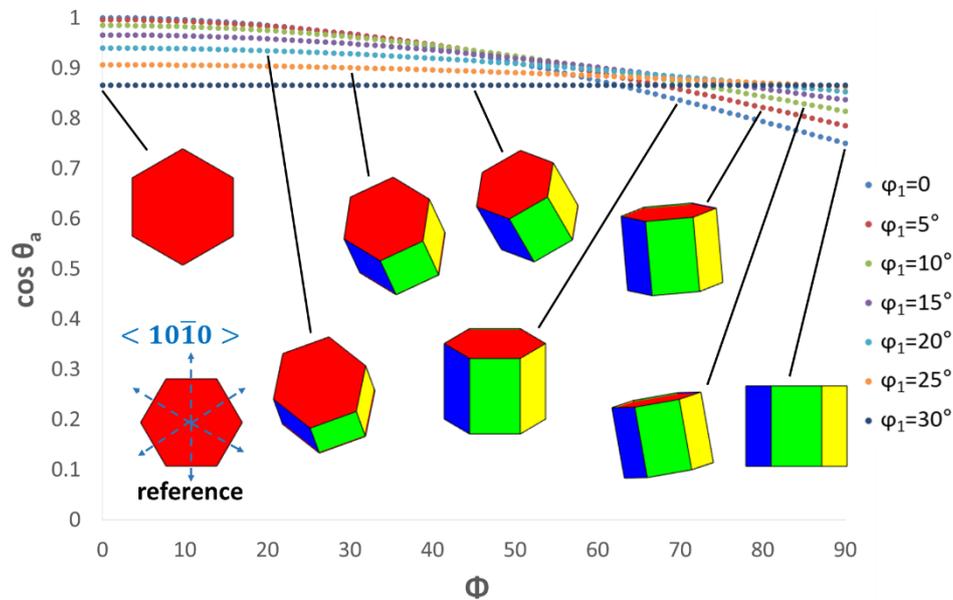

Figure 14 Change in the cosine of the $<1\bar{1}00>$ misorientation angle with $\varphi_1$ and $\phi$, using [0,0,30°] as a reference.

To explore the reason for the preferred orientation of the hydride stringers in the fine grain material, the grain boundaries where hydrides formed in Figure 9 were analysed in Figure 15, in terms of the angle between the surface traces of the grain boundaries and the basal planes of the grains where the intergranular hydrides formed (φ). Grains smaller than 40 µm² in the map were ruled out from the analysis as their boundaries likely make a relatively substantial angle with the surface traces. Results indicate a weak correlation between hydride precipitation and the φ value, and the preference of low φ grain boundaries for hydride nucleation as hypothesised in literatures [22,32] cannot be proved by the experimental results here.

In Figure 9, the texture of the material is such that the $\{10\bar{1}7\}$ planes of the grains are mostly close to the vertical direction. It is hence speculated that the *type-a* intergranular hydrides tend to form in the grains with their $\{10\bar{1}7\}$ planes aligned close to the grain boundary, while the morphology of *type-b* hydrides (thin and not along the grain boundaries) are due to their habit planes being relatively far from the grain boundaries. Therefore, in the fine grain material, the macroscopically observed preferential alignment of the hydride stringers perpendicular to ND is likely a collective effect of the thick morphology of the *type-a* intergranular hydrides and the fact that the $\{10\bar{1}7\}$ habit plane of the *type-b* intergranular hydrides being nearly perpendicular to the ND of the textured Zircaloy-4 plate.

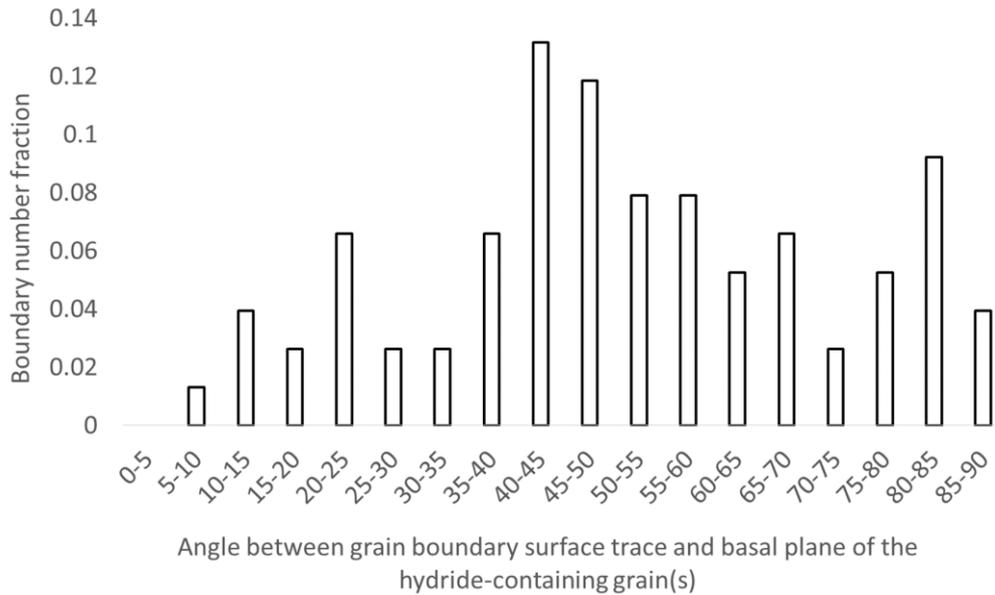

Figure 15 Distribution of hydride-containing grain boundaries with the angle between the grain boundary surface trace and the basal plane of the hydride-containing grain(s).

## 5. Conclusions

Various types of δ phase hydrides in textured as-received fine grain (~11 μm in average grain size) and 'blocky alpha' large grain (average grain size >200 μm) Zircaloy-4 have been investigated in detail using EBSD. Hydrides were formed using a moderate cooling rate (0.5 °C/min) and in these materials the following conclusions can be drawn.

1. All the hydrides examined in Zircaloy-4 with both grain sizes follow the $\{0001\}_\alpha||\{111\}_\delta; <11\bar{2}0>_\alpha || <110>_\delta$ OR with the surrounding matrix.
2. The macroscopic habit planes of the intragranular hydrides and the intergranular hydrides which nucleated at the grain boundaries and grew into the grain interior are the $\{10\bar{1}7\}$ planes of the surrounding matrix.
3. The $\{10\bar{1}2\}<\bar{1}011>$ T1 twin boundaries in 'blocky alpha' Zircaloy-4 are preferential nucleation sites for hydrides, and the macroscopic habit planes of the hydrides at T1 twin boundaries are parallel to the $\{10\bar{1}2\}$ twin plane.
4. The morphology of the intergranular hydrides depends strongly on the angle between the grain boundary and the $\{10\bar{1}7\}$ planes. For grain boundaries aligned close to the $\{10\bar{1}7\}$ planes, relatively thick hydrides are formed along the boundary. For other boundaries, thin hydrides are formed which preferentially decorate the $\{10\bar{1}7\}$ planes leading from the boundary and into the grain, instead of forming along the grain boundary.
5. The misfit strain between α-Zr and δ-hydride is accommodated by dislocations in both the matrix and the hydrides and twinning in the hydrides. The dislocation density is generally higher in the hydrides than in the matrix, and the twinning in the hydrides can be caused by the changes in the Burgers vectors of the $\frac{a}{3}<1\bar{1}00>$ partial dislocations upon hydride formation.

6. The growth of hydrides across grain boundaries is achieved through an auto-catalytic nucleation process, that is, the modulation of the local chemical potential distribution in the grain ahead of the hydride growth front due to the existing hydride leads to the hydride nucleation on the other side of the grain boundary.
7. Intergranular hydride nucleation favours grain boundaries with low (<40°) and high (>80°) *c*-axis misorientation angles. Presumably, the former is due to easy collective shear on basal planes required for the transformation and the latter is due to high grain boundary energy.
8. Intergranular hydride nucleation does not prefer grain boundaries close to the basal planes of the adjacent grains. The alignment of the hydride stringer observed macroscopically is due to the alignment of the $\{10\bar{1}7\}$ habit planes perpendicular to ND as a result of the bulk material texture.

## Acknowledgements

The authors acknowledge Dr Vivian Tong for useful discussions and developing the Matlab code for grain boundary misorientation analysis. TBB and SW acknowledge support from the HexMat programme grant (EP/K034332/1). TBB thanks the Royal Academy of Engineering for funding his Research Fellowship. The FEI Quanta SEM used was supported by the Shell AIMS UTC and is housed in the Harvey Flower EM suite at Imperial College London.